# Coupled plasmon wave dynamics beyond anomalous reflection: phase gradient Copper metasurface for visible to the near infrared spectrum.

HOSNA SULTANA,[1,2,*]

1. Electrical and Computer Engineering, The University of Alabama, Tuscaloosa, AL 35487, USA.
2. Remote Sensing Center, The University of Alabama, Tuscaloosa, AL 35487, USA.

**Abstract:** In nanoscale photonic devices, the demand for multifunctionality from metasurface 2D optics increases rapidly. To explore fine-tuning in the design metric, we reinvestigated the trapezoid shape copper metasurface with Finite-Difference Time-Domain simulation for efficiently using linearly polarized light for two different functionalities. From the plasmonic band structure, we see how the degree of asymmetry in geometry affects the efficient resonance coupling of the traveling plasmonic modes, along with different types of mode hybridization profiles related to the nanoantenna's geometric shape. Tuning the nanoantenna's length, we can excite the effective plasmon mode supported by this configuration and guide out surface wave unidirectionally from the normal incident free-space light for visible to infrared range. Directed surface plasmon polariton has both antisymmetric and symmetric modes oscillating between the top and bottom surface of the continuous metal layer depending on the nanoantenna's length and wavelength. This proposed Copper metasurface is optimized for far-field application of broadband (600-900 nm) anomalous beam steering for an average of 65 percent efficiency with a maximum of 64 degree angle. This work brings more understanding of how one metasurface can be implemented in small footprint plasmonic devices, waveguide mode controlling, as well as beam steering with their wavelength dependent functionality.

1. Introduction

The demand for subwavelength scale engineering for compact, flexible optics and exploring the critical limit of light-matter interaction is proliferating. The promising application of these optoelectronic devices for optical holography, superlens imaging, polarimetric detecting, beam steering, waveguide coupling, beam splitting, and vortex beam generation successfully replaces bulky optics [1–4]. Though remarkable results and utility were demonstrated with optical metasurface for the photonic device, the multifunctionality and wavelength depended multiplexing remain challenging [5,6]. The plasmonic metasurfaces have become more popular because of their subwavelength scale aspect ratio, and Gap Surface Plasmon (GSP) assisted mode tuning option to get better control over resonance loss, so high-efficient controlled reflection orders with the magnetic coupling [5,7–9]. Metasurface with Ag trapezoid shape nanoantenna has been investigated by Li Z. et al. for visible wavelength to demonstrate the angle-resolved broadband beam splitting [5]. Concurrent reporting from Zhang Y. et el. came with Au trapezoid [10]. Z. Lei et al. came up with the new idea of tuning the resonance coupling between the dipole and their imperfect image of Ag trapezoid to achieve both anomalous reflection and transmission at the visible wavelength, but fabrication-wise the design is challenging [7]. Gao S. et el. did excellent work of wavelength insensitive phase gradient trapezoid nanoantenna with Al for propagating wave (PW), but without exploring more on the plasmonic effects [11]. Sun et al.'s work with discrete bar nanoantenna connects phase gradient to the onset of the surface wave (SW) and the angular dependency of anomalously reflected beam [12,13]. Li. Z et el. demonstrated the mode conversion, polarization rotation, and the asymmetric coupling of the waveguide modes for mid IR range with a small footprint metasurface with gradient gold nanoantenna over the dielectric waveguide [14]. Within one decade, many works have been reported with these approaches of new possibilities, yet more investigation needs to be done on how one metasurface can be used efficiently for both PW and SW in a controlled manner.

In this article, we have reported the Cu metasurface for the plasmonic material. Both Au and Ag have been extensively investigated in literature but not that much for Cu. The interband transition energy level for Cu is 2.15 eV and has the capability of interband excitation at visible wavelength as compared to Au, which case is 4 eV [15]. The oxidation of Cu is an issue, which in recent years has been improved by capping agents [16,17]. So, the promising photonics device can be made cost-effectively with Cu. With Finite-Difference Time-Domain (FDTD) simulation (FDTD Solutions, Ansys/Lumerical), we explored the design of the metasurface optimized for visible to the infrared regime for anomalous reflection and SPP channeling. This article is divided into two sections to discuss these two aspects of our metasurface design.

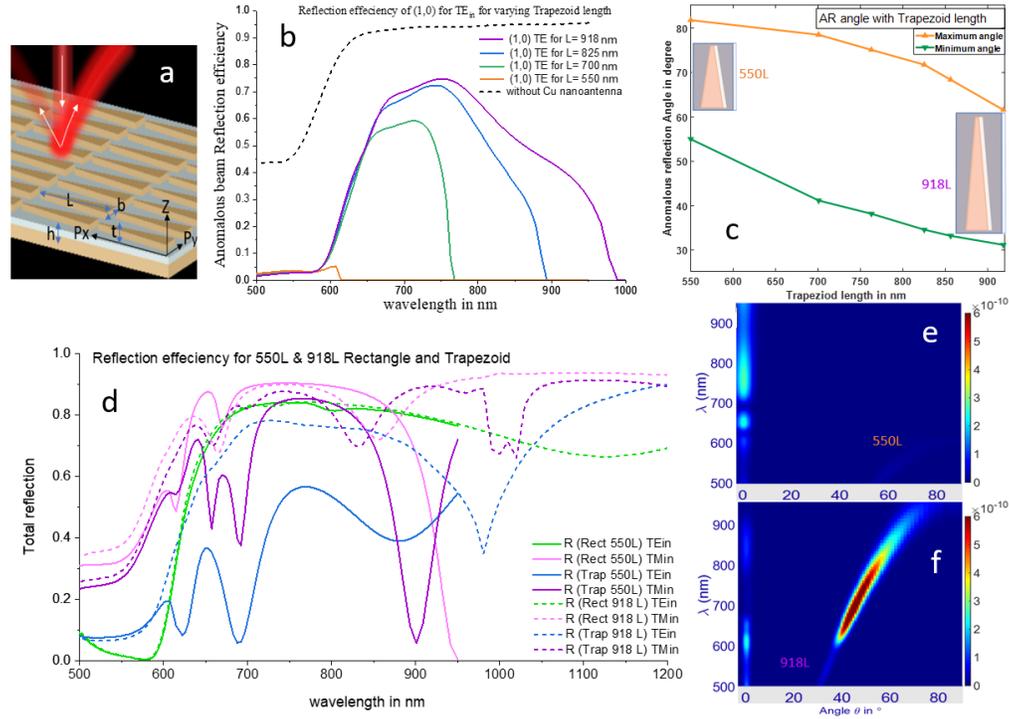

**Figure 1**. (a) Cu-Metasurface schematics. (b) Broadband anomalous reflection efficiency for four different length trapezoids as mentioned in Table S1 of the supporting information text. (c) Minimum and maximum angle coverage by the anomalously reflected beam. (d) Comparison of the broadband total reflection efficiency for 918L (dashed line) and 550L (solid line) trapezoids and rectangles for both TE and TM incidence. Angle-resolved far-field reflection magnitude depending on the wavelength for TE incidence for (e) 550L trapezoid, (f) 918L trapezoid.

2. **Simulation and modeling:**

We choose the thickness of Cu trapezoid nanoantenna 50 nm ($\lambda/10$ - $\lambda/24$) to overcome skin depth for Cu-SiO$_2$ system over a continuous SiO$_2$ spacer (h = 50 nm) and a 100 nm thick Cu bottom layer (figure 1a), for the working wavelength 500-1200 nm. The value of the optical constant for simulation and calculation is taken from palik [18]. We keep a constant gap of 62 nm between one tip to another base and constant base width b = 185 nm, tip = 20nm, and periodicity toward base $P_y$ = 349 nm. We changed the length (L= 918, 825, 700, 550 nm) as well as corresponding periodicity $P_x$ towards length direction (table S1 in supplementary information) to optimize the lengths of the trapezoid. Plane wave source is used for the reflection, phase profile, and wavelength dependent SPP excitation with a Bloch boundary condition in x and y with a PML towards z. For scattering and absorption cross-section, power flow, and SPP wave propagation, a Total-Field Scattered-Field (TFSF) source has been used from the top with all PML boundaries. For the band structure calculations, modes are excited with the randomly oriented electric dipole near Cu-nanoantenna/$SiO_2$ interface and obtain the result up to the first Brillouin zone edge. In this article, our main interest is for 918 and 550 length trapezoids, and we will frequently compare them by 918L and 550L names.

3. **Results and discussion:**

The phase shift offered by the trapezoid nanoantenna, which is symmetric under incident y-polarized light, varies linearly with the x-position, conveyed with dφ(x)/dx in the generalized form of Snell's law:

$$n_{r,t}\sin\theta_{r,t} - n_i \sin\theta_i = m_0 \frac{\lambda_0}{P_x} + \frac{\lambda_0}{2\pi}\frac{d\Phi(x)}{dx} \quad (1)$$

Where $n_{r,t}$ and $n_i$ are the refractive index of the reflected and transmitted medium, respectively. The subscript with m signifies the diffraction order. Equation 1 is a simple grating equation with a maximum of 2π accumulation of the additional phase within the one periodicity of the unit cell; the term gives $d\varphi(x)/dx = 2\pi/P_x$

$$n_{r,t}\sin\theta_{r,t} - n_i \sin\theta_i = (m_0 + 1)\frac{\lambda_0}{P_x} \quad (2)$$

So, the prominent order of power flow shifts to the positive 1st diffraction order after reflection or transmission [5,7,19]. For wider angle high-efficiency anomalous reflection for 500-950 nm broadband regime, we have come up with the optimized aspect ratio and length of the trapezoid shown in figures 1b, 1e, 1f, and S1.1b. Below 550 nm wavelength, the reflection efficiency of plain $SiO_2$/Cu surface is low. So, Cu is not a good candidate, but above that wavelength, the efficiency can be raised to almost 70% as the length of the trapezoid increases with the compensation of the AR reflection angle. Figure S.1.2 (a) in the supporting information section provides all the simulated wavelength-dependent angle-resolved far-field anomalous reflection magnitude for all the trapezoids, both for TE ($E_{in} \parallel y$) and TM ($E_{in} \parallel x$). Also, the simulated wavelength-dependent total reflection efficiency and the specular and (-1,0) order reflection efficiencies are presented in figures S1.2b and S1.2c. Figure 1d shows broadband total reflection efficiency bringing the rectangle and trapezoid in comparison for this two-length size, both for TE and TM normal incidence. If we look at the solid and dotted green lines, we see they almost fall in line as the E-field in TE incidence does not experience any difference in the geometry change. In this case, TM incident shows sharp resonance dips related to the periodicity and gap surface plasmon resonance. Interestingly for the 550L rectangle, the resonance phase shift is huge to suppress the reflection totally around 950 nm of wavelength. Magnetic dipole resonance modulates the phase of the reflected beam. Among the trapezoids for the TE incidence, the strength of the dipole resonance enhances in the localized region at different widths with the wavelength, as shown in figure 2b. [2,5,8,11,14,20]. The beam steering limit for tor these trapezoids is presented in figure 1c, plotting the AR angle minimum and maximum. The angular limit for the 1st order diffraction for our design is $0.537 < \lambda_0/P_x < 1$ for the 500-950 nm spectral band.

The phase profile is presented in figures 2a and 2b. The increasing width of a bar-nanoantenna can be represented as the trapezoid's increment length position [5,11]. The irregularity in the phase contour in the y-phase profile starts to build up from the trapezoid's base side mostly (figure 2a), from about 70 nm of the width towards the base: for both cases of 550L and 918L, around 611 nm of wavelength. Anomalous reflection starts to become dominant around this wavelength (figure 1a and S1.2a). In the wavelength-dependent broadband reflection phases, as shown in figure 2b, we can see that for the 550L trapezoid, the peculiar phase hold from 622-633 nm of wavelength is because of the shifting of the dipole resonance position along the width, from the the middle towards to the base side. Within 634-657 nm of wavelength, the resonance enhances strongly in the same position. This does not occur for the rectangle nanoantenna shape we see in figures 3b. If we compare x-phase (figure 2b) with the TM reflection profile (figure 1d), for the 550L trapezoid, the abrupt phase change around 900 nm of wavelength causes the reflection to drop, and that can be attributed to the cavity gap surface plasmon (GSP) resonance, as we see when bringing comparison with the rectangle [8,10,19,21]. Same for the 918L trapezoid; the small amount of phase change at 655 and 809 nm of wavelength shows the effect of reflection drop. For TM incidence, when the $sin\theta_{r,t}$ for Eq. 2 becomes imaginary at $\lambda_0 > P_x$, the surface-bound evanesce mode originates. The in-plane component of the field interacts with metal's free election and form of collective oscillation, which remains confined at the dielectric metal interfaces. We will discuss plasmonic mode excitation in the 2$^{nd}$ section. We can calculate the phase gradient for any wavelength from the phase color map. The amount $d\varphi_x/dx = 2\times10^{-3}$ rad/nm for 550L trapezoid and $d\varphi_x/dx = 7\times10^{-4}$ rad/nm 918L trapezoid at 637 nm and 990 nm of wavelength respectively under normal incident. In the next section and supplementary text section 2, we will see why the phase gradient at the 637 nm and 990 nm of wavelength are of our interest.

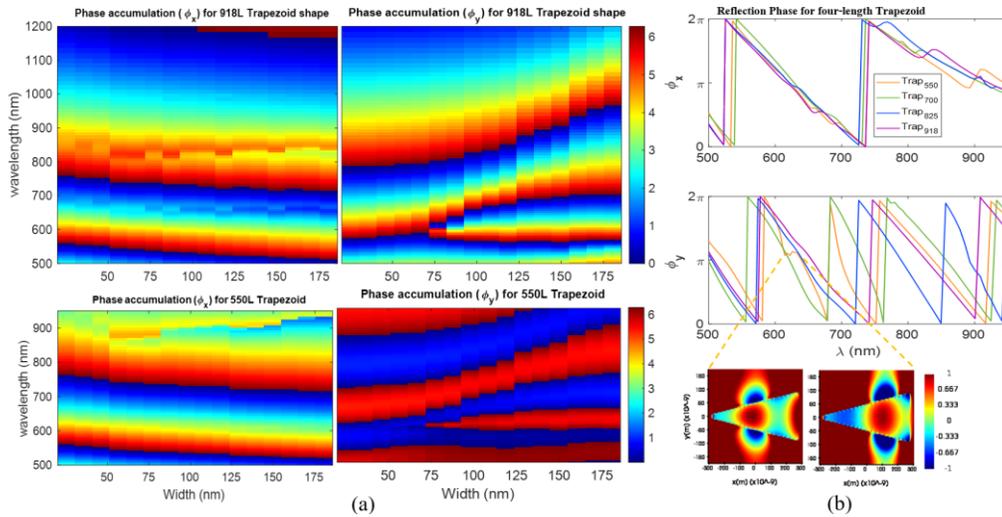

**Figure 2.** Phase gradient in effect from trapezoid shape gradient: (a) Simulated 2D phase map for Cu-bar nanoantenna of increasing width with wavelength for 918L (Top) and 550L (bottom), respectively. Color bar scale value is in radian. (b) Wavelength dependent TM and TE phase for four different lengths Cu-trapezoid, where the abrupt changes occur by shape-dependent resonance modes for 550L trapezoid but do not affect the 918L trapezoid.

Thanks to the phase gradient metasurface, which can couple light to the SPP mode from a normal incidence beam with a flat two-dimensional geometry [22–25]. The detailed mechanisms are explained in supporting information section S2.1. In brief, from figure 3a, we can explore the incidence angle-dependent reflection profile for 550L trapezoid for the wavelength range 600-700 nm, where the phase profile gets interesting. At 637 nm of wavelength, the reflection drop indicates coupling to plasmon modes at normal incidence and 12$^0$ angles as reflection drop almost the same magnitude. Figure 3b compares wavelength-dependent phase profiles for rectangle and trapezoid geometry. Figure 3c and 3d show $\mathbf{E_z}$ field profiles indicating SPP mode excitation effectively for trapezoid shape. And while we compare the $\mathbf{E}_z(z,x)$ for various length positions for both (figures 3c and 3d), we can clearly see the magnitude and nature of the field excitation. For the trapezoid, almost twice the field enhancement to tip direction compared to base and dynamically changing the interplay between the symmetric and antisymmetric modes of the transverse field vector between Cu-nanoantenna and the bottom Cu layer originates the unidirectional SPP propagation (visualization 1). For the rectangle, the transverse field vector's antisymmetric modes between two metal regions flip sign from one edge to another, indicating the rectangle works as a perfect dipole with an almost standing SPP wave (visualization 2).

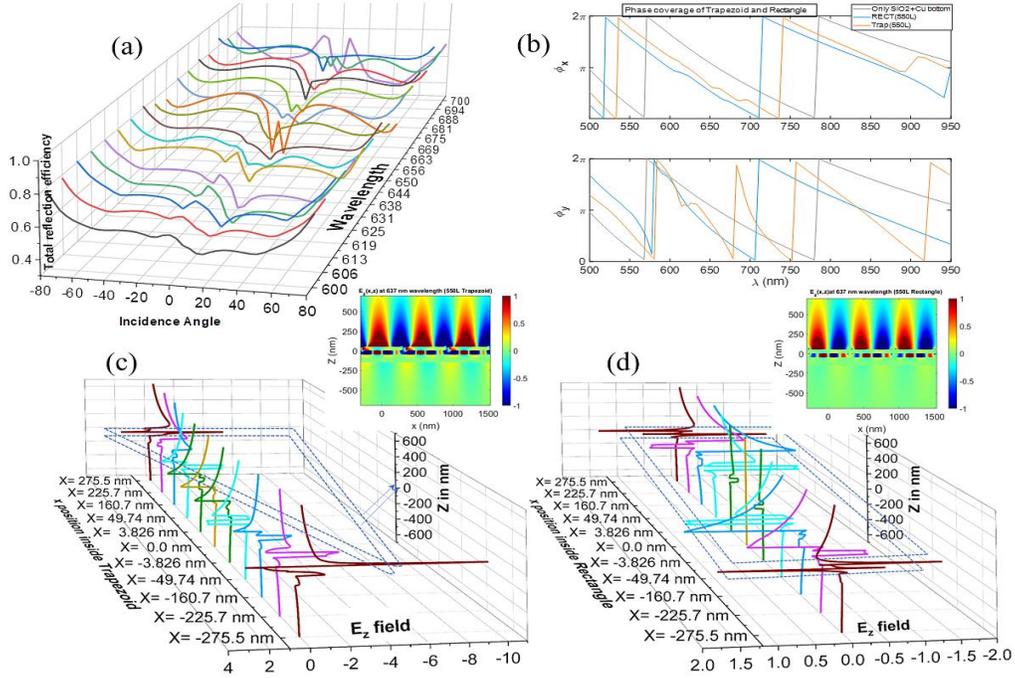

**Figure 3.** (a) Simulated TM incident angle dependent reflection efficiency for 550L trapezoid in the wavelength range 600 to 700 nm to see the normal incidence origin of SPP. (b) Comparison of wavelength dependent phase profile between 550L trapezoid and rectangle. $E_z$ (z) field profile at various x-positions inside the (c) trapezoid and (d) rectangle. Here trapezoid is from z=0 to 50 nm, the continuous $SiO_2$ layer occupies Z=-50 to 0 nm, and the Cu bottom is from z=-150 to -50 nm. The blue dash line shows the structural outline. The inset shows the $E_z(x,z)$ field profile for the three trapezoids (c) and rectangles (d).

The SPP dispersion relation for both air/Cu and $SiO_2$/Cu interface for the metasurface continuous layers is given in figures 4a and S2.1a. The huge difference in the theoretical SPP propagation length in terms of wavelength for air/Cu and $SiO_2$/Cu interface along with SPP wavevector is shown in figures S2.1 b and c. Theoretically, the SPP propagation length for Cu/$SiO_2$ interface is 1.77 μm, and 3.8 μm for 637 nm and 990 nm of $\lambda_0$ and for the Cu/air interface is 11.35 μm and 270 μm for 637 nm and 990 nm of $\lambda_0$ respectively. Figures S2.2 shows the SPP excitation profile with $E(x,z)$ and $E_z(x,z)$ of these structures for the three trapezoids in a raw with a comparison for rectangle and trapezoid. Section S2.3 depicts the wavelength dependent $E_z(z,\lambda)$ and magnitude $E(z,\lambda)$ profile which validates the dispersion curve and how these two specific length trapezoid's geometries are effective in exciting plasmon modes for two specific wavelengths.

We check the band structure for the metasurface to know which wave vector contributes to any photonic states, as presented in figure 4. The color bar indicates the coupling strength. From the dispersion relation shown in figure 4a for the continuous layer interface, we can see the system support two SPP modes. At the low-frequency regime, the SPP and light lines are alike, then around 303 THz (990 nm of $\lambda_0$) and 470 THz (637 nm of $\lambda_0$) SPP modes branch off on

Cu/SiO$_2$ interface and at Cu/air interface, respectively. When the Cu-nanoantenna is present, the effect phase gradient creates complicated GSP mode behavior. For broadband ($\lambda_0$= 500-1200 nm), the frequency range 250-600 is of our interest. Two light lines that go straight to the Brillouin zone boundary are from SiO$_2$ and air.

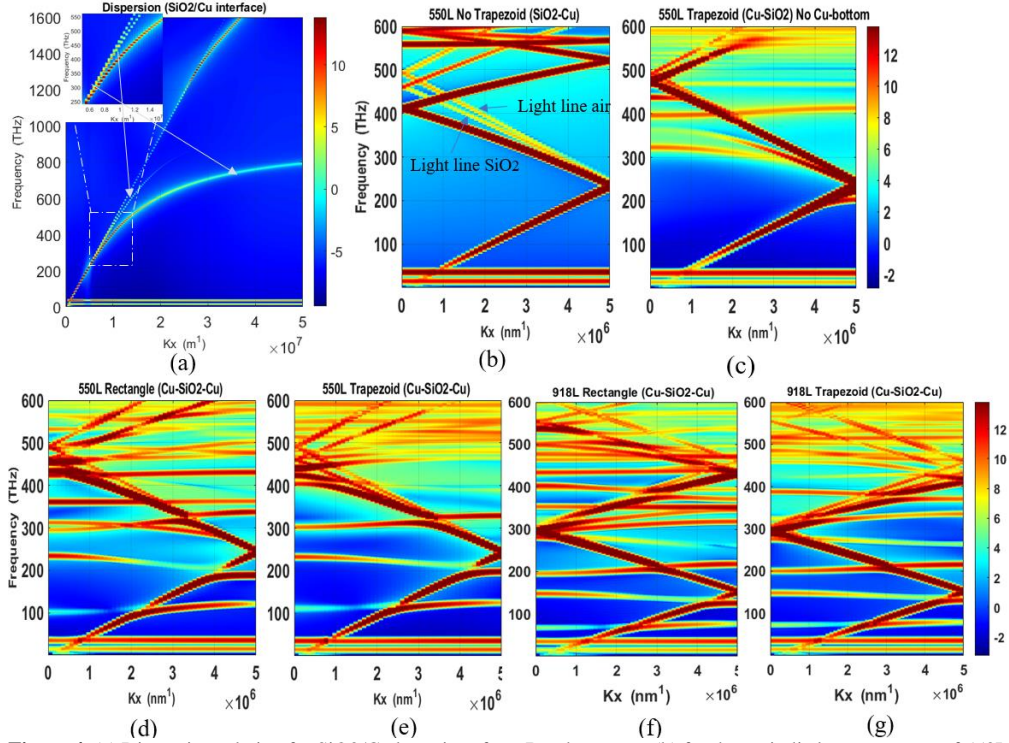

**Figure 4.** (a) Dispersion relation for SiO2/Cu layer interface. Band structure (b) for the periodic layer structure of 550L trapezoid but without the Cu trapezoid, (c) for the periodic structure of 550L trapezoid but without the Cu bottom, (d) and (e) for the periodic structure of 550L rectangle and trapezoid. (f) and (g) for the periodic structure of 918L rectangle and trapezoid.

To compare the simplest case of the dispersion curve folded in the first Brillouin zone, we calculated the band structure for 550L trapezoid's periodicity, as shown in figure 4b but without the trapezoid. Figure 4c shows band folding for the structure of the 550L trapezoid and SiO2 layer but without Cu bottom layer. In the comparison of figures 4b and 4c, it is clear the five new bands belong to the Cu trapezoid. Without any bottom Cu layer coupling, the dispersion reaches the zone boundary at 206 THz. These modes below the light line have the ability of strong light coupling to produce localized optical modes [26–32]. The appearance of more of the resonant modes at the zone center for of the presence of the nanoantenna indicates the origination of SPP branches and their traveling status when they have a non-zero slope. Comparing figures 4b, 4c, 4d, and 4e, we can see the plasmonic mode hybridizing profile and the bandgap opening when all the structure components (trapezoid, SiO$_2$, and Cu-bottom layer) are present [30,33]. The bandgap opening at $K_x=0$, has an analogy with the charge density distribution, where the solution of the different energy long-range surface plasmon polariton (LRSPP) and the short-range surface plasmon polariton (SRSPP) has self-crossing [34]. The 550L rectangle and trapezoid comparison yield two extra almost localized modes appearing for rectangle around 320 and 360 THz. Suppose we compare the resonance dip of the reflection curve of figure 1d with a normal incidence ($K_x=0$) mode, the strong coupling 438-453 THz for the 550L rectangle is responsible for the reflection drop around 669 nm of wavelength and coupling of 490 THz is for the dip of 614 nm wavelength and coupling around 315 THz is for total reflection loss 950 nm of wavelength. Also, for the 550L trapezoid, the plasmonic mode coupling at 490, 456, and 436 THz accounted for the resonance dip at 614, 657, 687 nm of wavelength, but the trapezoid has more diffractive localized states at higher frequencies as a base and tips couple light differently. Since group velocity $v_g = d\omega/dk$ is equivalent to the slope of the dispersion curve, the localized states can be seen near the resonance frequencies where the slope tends to go to zero, indicating the large local density of states. But without the momentum match, these modes cannot couple light from the free space. So, having the traveling GSP resonance modes are important in the mechanism in order to couple light from free space [28–30]. Figures 4f and 4g and brings 918L rectangle and trapezoid in comparison show more localized GSP bands from standing plasmon modes for the rectangle. This scenario may be responsible for why the 918L rectangle mostly does not excite SPP on the Cu-bottom layer as there is less coupling of modes between top nanoantenna to bottom layer through the $SiO_2$ layer [28,35]. Gap plasmon are localized in the dielectric layer, not the gap between the Cu-nanoantenna's (figures S2.2b).

The effect of the GSP resonance can be seen in the scattering and absorption cross-section [8,32]. The absorption peaks around 809 nm wavelength for 918L and 834 nm of wavelength for 550L trapezoid are related to the power conversion to the surface wave, which propagates to the opposite direction almost equal magnitude canceling net effect. We can compare the optical power propagation for the scattering and surface wave counterparts in figure 5b to estimate the ratio of power converting to surface wave. For the trapezoids, the net optical power flow is towards the tip direction for the entire broadband range indicated (solid cyan line) for 550L and (solid green line) for 918L. The nature of this guided surface wave is related to their plasmonic counterpart through metal-insulator-metal cavity, especially when they are unidirectional preferentially due to phase gradient metallic nanoantenna. Waveguide mode coupling using phase gradient metasurface has been reported in the literature, where asymmetric optical power transmission mode conversion happens on an oppositely propagating beam in the cavity [14,21,24,36,37].

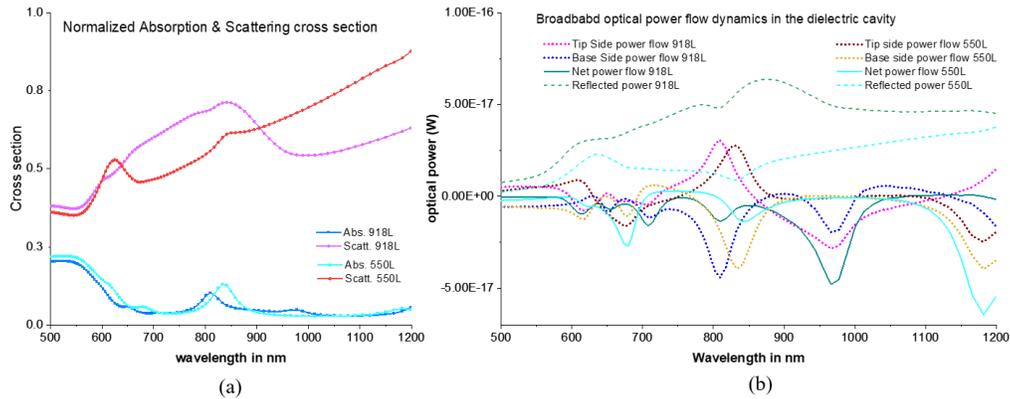

**Figure 5.** (a) Normalized scattering and absorption cross-section. (b) power flow of SPP wave. Here the negative value indicates power is flowing towards the -x direction (tip direction). Power flowing towards the +x direction (base direction) is showing positive value as well as the reflected power (+z direction).

When the incident waveguide mode propagates towards the phase gradient direction of the metasurface, the wave vector increases continuously, and the optical power gets coupled to propagating surface wave. When waveguide mode propagates opposite to phase gradient, the wave vector bending angle increases. Both processes involve waveguide mode conversion oppositely to a lower and higher order, respectively [14,38,39]. SPP wave propagation for both trapezoids shown in figure 6. Here four 550L trapezoids are illuminated at 637nm of wavelength (figures 7a,7c,7e) and three 918 trapezoids by 990 nm of wavelength (figures

7b,7d,7f). More details are given in the supporting information text section S3 and the supplementary visualization 3, visualization 4 for 550L and visualization 6, visualization 7 for 918L trapezoid, respectively, for E and $E_z$ components. As we see, the SPP wave launches from the metal backplane under the trapezoid region, and long-range collective coupling between the top surface (Cu/SiO$_2$) and the bottom surface (Cu/air) of the continuous Cu layer results in the propagation of transverse unidirectional SPP wave prominently toward to tip direction for this wavelength. For the rectangle, the field excitation to SP mode dissipated and scattered without significant guiding out of the SPP wave shown in visualization 5 and visualization 8 for 550L and 918L rectangle, respectively. From the band structure, we see the rectangle has both forward and backward propagating modes, whereas the trapezoid has mostly forward propagating modes [40]. The $\lambda_{spp}$ is approximately 610 nm for 550L trapezoid and 880 nm for 918L trapezoid as we can determine from figure 6e. A similar type of SPP wave propagation has been reported in the literature where the in-phase or out of phase relation of corrugation pattern on the two surfaces of a thin metal layer can link the LRSPP and SRSPP [34,41,42]. Interestingly, the E-field and vector profile (figure S2.2c) between Cu nanoantenna and under the Cu bottom layer matches the similar description [34]. In our metasurface, the bottom layer does not have any corrugation pattern or structure, so in that regard, the oscillatory coupled SPP propagation in two plain surfaces of the Cu bottom layer is worthwhile to investigate more. Figures 6e and 6f reveal the symmetric (for 550L) and antisymmetric (for 918L) plasmonic modes for the respective wavelengths [40,43]. The coupling continuous even greater extent through the Cu bottom layer up to 500 nm of the thickness of as shown in figure S.3 indicates antisymmetric modes of LRSPP prevail. The wavelength of this coupled guided oscillation can be tuned by varying the Cu-bottom layer thickness as the trend shows longer oscillating wave of LRSPP for thicker Cu-bottom layer. For SRSPP, also the guided oscillation wavelength for the half wave on the Cu-bottom layer top surface become smaller loosing coupling with bottom surface with increasing Cu-bottom layer thickness. For broadband, the collective propagating SPP wave is more dispersed due to more diffractive GSP modes launched with different propagation vectors that will be the topic for future investigation.

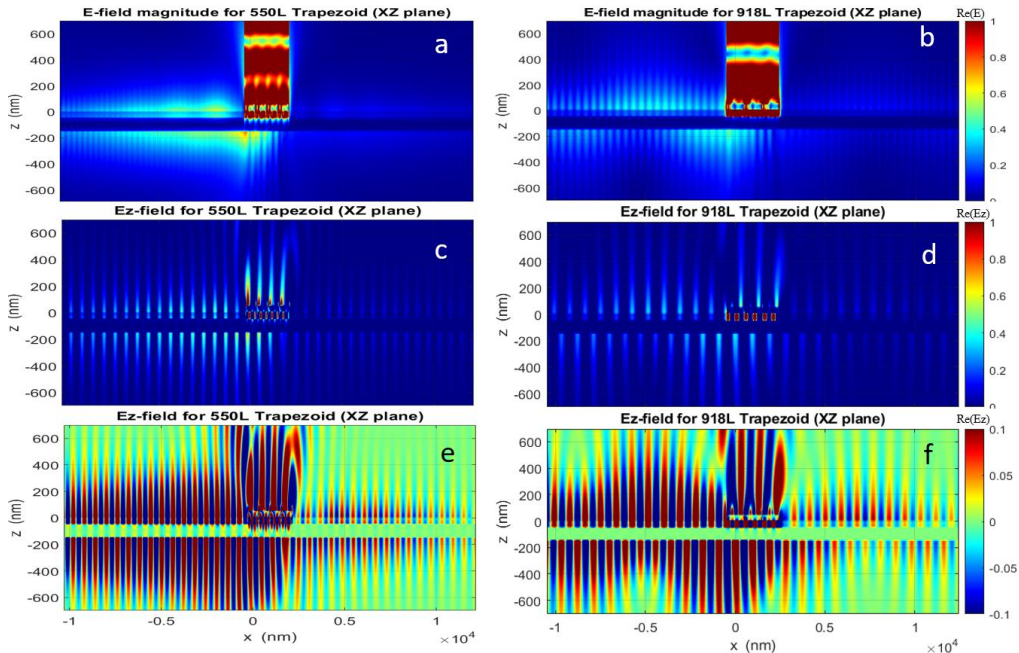

**Figure 6.** SPP channeling by symmetric and antisymmetric modes for up to 10 μm beyond the trapezoid area in both tip and base direction at (Y=0) XZ plane. (a),(c) & (e) are for 4 of 550L trapezoids illuminated at 637 nm of λ. (b), (d)

& (f) are for 3 of 918L trapezoids illuminated at 990 nm of λ. The bottom one enhances scaling the Ez component for showing symmetric SPP mode (for 550L) and antisymmetric SPP mode (for 918L) trapezoids through the $SiO_2$/Cu-bottom layer and Cu-bottom layer/air interfaces.

## 4. CONCLUSION

In conclusion, we proposed a simple phase gradient metasurface design and investigated the far-field scattering and plasmonic aspect of it in depth with FDTD simulation for utilization of both orthogonal linearly polarized light. We correlated the shape-dependent reflection property with phase profile and details scenario of anomalous beam steering. We verify the propagating coupled SPP wave with the same metasurface and explain the underlying phenomenon with band structure and scattering cross-section and field profile. This simple design geometry can accomplish efficient multifunctionality as an element in the solar sail, nano-photonic and optical communication devices, from visible to near-infrared bandwidth.

**Corresponding author:** hsultana@crimson.ua.edu

**Acknowledgement:** H. Sultana acknowledges the financial support from the Remote Sensing Center (RSC) at The University of Alabama, Tuscaloosa.

**Informed Consent Statemen:** The author declare no competing financial interest.

**Data availability:** Not applicable.

**Supplemental material.** See the supplement text and associated visualization content.